# Challenging Techno-Solutionism: The Role of ICT Innovation and the Value of Technological Growth

Srinjoy Mitra[1], Henrik Skaug Sætra[2], Jean-Pierre Raskin[3], Jan-Christoph Heilinger[4], Mario Pansera[5]
[1]University of Edinburgh, UK; [2]University of Oslo, Norway; [3]Université Catholique de Louvain, Belgium; [4]Witten/Herdecke University, Germany; [5]Universidade de Vigo, Spain

***Abstract*— Innovation in Information and Communication Technology (ICT) has become one of the key economic drivers of our technology-dependent world. Digital devices/systems have become so pervasive that it is hard to imagine new technology developments that are not totally or partially influenced by ICT innovations. Furthermore, the pace of innovation in ICT sector over the last few decades has been unprecedented in human history. In this paper, we argue that the ICT innovation paradigm has crucially shaped collective expectations and imagination about what technology more broadly can actually deliver, particularly for a more sustainable and equitable world. These expectations have often crystalised into a widespread acceptance, among general public and policy makers, of techno-solutionism. We emphasise the role of electronic microchips in deriving relentless innovation and its impacts. We identify the many impacts of this innovation cycle into three different categories: embodied, induced and implied impact. We show what drives this innovation mania (microchips and Moore's Law) and how this impact categorisation helps us understand its long-term repercussions (from material consumption to ideological dominance). The spectacular advances in this sector, coupled with corporate power that benefits from them, have facilitated an uncritical narrative of techno-optimism that neglects the complexity of the 'wicked problems' that affect the present and future of humanity. We conclude with suggestions that the innovation paradigm has to be reconsidered and limits to ICT growth have to be introduced.**

## 1. Introduction

The world has seen unprecedented innovation in ICT for over five decades. There is little doubt that ICT is the most prominent general-purpose technology of today (Kooij, 2023). It is the core of everything 'tech', from computers in our pockets to satellites in space. ICT innovations have affected modern lives in innumerable ways and are constantly creating many opportunities that were impossible to even dream of a few years ago (Liao et al., 2016). The consequence of the spectacular growth of ICT is that there is no end to this innovation pipeline. In fact, it seems the common consensus is that further growth in ICT is required to solve all sorts of problems, including environmental challenges, as witnessed in, for example, endeavours to push the *twin transition* (Muench et al., 2022). This indicates how digital technologies are seen not as contributors to environmental harms and a part of the problem, but instead as the necessary *solution*. This idea is emotively captured by an influential Silicon Valley financer/thinker, Marc Andreessen, in his "techno-optimist manifesto", which pushes this narrative to its extreme (Andreessen, 2023). More powerful ICT innovation, Andreessen, Bill Gates and others argue, will be our salvation and solution (McAfee, 2019). Enormously profitable companies are, however, waiting to reap ever larger profits by keeping this innovation cycle going. There are around ten trillion-dollar companies on earth, and almost all of them belong to the ICT industry (Ventura, 2025). These, and several other similar ICT companies, spend way more on research and development than the combined Western governments and publicly funded institutes ever could (Mitra & Raskin, 2023).

At the same time, there is abundant evidence of negative environmental impact in creating, using and disposing of these technologies (Santarius et al., 2023; Zhang & Wei, 2022). As Artificial Intelligence (AI) tools became one of the most prominent outputs of this sector, significant scholarship has gone into unearthing their inherent racial bias (Benjamin, 2019), the creation of surveillance states (Zuboff, 2019) or the overall impact on marginalised societies (Crawford, 2022).

Apart from these obvious impacts, the relentless cycle of ICT innovation also fostered the idea that technology is the solution to all problems, past and present. The idea that technology will, overall, improve the human condition is what is often referred to as techno-optimism (Danaher, 2022). Beyond optimism, however, we explore the more radical notion that technology will not only generally improve things, but that what is referred to as *wicked problems* – hunger, injustice, war, discrimination, poverty, climate change etc. – can also be *solved* with technological fixes or solutions (Johnston, 2020; Weinberg, 1966). Such so-called *techno-solutionism* entails a faith in and preference for technological solutions (Morozov, 2013; Sætra & Selinger, 2023). Instead of seeking to address environmental challenges and inequality through social change, the techno-solutionists mostly think that these challenges can be solved with more and better technology. This ideology enables a belief in new technology looming on the horizon, with little need to adjust our existing (extractive and unjust) practices. What could be better?

The problem is that such techno-fixes and techno-solutionism are prone to be accompanied by a range of negative, unanticipated consequences, and they often even fail to address the social problems they were argued to solve. As Acemoglu and Johnson argue, technological innovations/solutions alone almost never result in shared prosperity and thus must not be conflated with progress (Johnson & Acemoglu, 2023). Similarly, scholars have argued that the idea of technological innovation as always good and desirable should be reconsidered (Coad et al., 2021; Vinsel & Russell, 2020). Furthermore, within this 'innovation mania' there are multiple implicit beliefs about social prosperity that leave no space for 'alternative socio-technological imaginaries about growth and development'(Pansera & Fressoli, 2021).

In this paper, we suggest that continuous ICT innovation, particularly in electronic microchips, is one of the biggest drivers of techno-solutionism and an overly optimistic belief in a techno-utopia. This belief has become increasingly pervasive within the general public and policymakers, particularly as a way to counter climate change. Even progressive thinkers and politicians supporting the US Green New Deal or European Green Deal, are not immune to the techno-solutionism trap, an "unproven premise that technology can undo the harm done without causing further harm, either in the farther future or in the form of externalities" (Ossewaarde & Ossewaarde-Lowtoo, 2020).

In what follows, we analyse these developments not only empirically but also normatively. Our assessment is guided by a sustainability- and a justice-based framework: Technological innovation should be evaluated by its contribution to maintaining ecological thresholds and enabling equitable social outcomes. From this standpoint, the dominant ICT paradigm fails because it externalises ecological costs and consolidates and reinforces inequalities.

The paper is structured as follows. The following sections show that ICT can be a proxy for all modern technology (section 2) and has dominated the global industry since the 1970s (section 3). Section 4 analyses the different forms of morally problematic impact of ongoing ICT innovation in three dimensions, which we call their embodied, induced, and implied impact. Section 5 exposes the limits of an optimistic belief in ICT techno-solutionism, making a case for limiting and reining it in. Section 6 concludes.

## 2. ICT as a proxy for all modern technology

Information and Communication Technology (ICT) is a catch-all term for a wide variety of devices and tools. This goes beyond computation/communication devices (e.g., computers, phones, data centres) to encompass a wider set of technological systems and artefacts. Popular magazines focused on consumer gadgets have helped blur the boundary between 'technology' and ICT technology even further, particularly when we talk about technological innovation. However, there is some grain of truth to this. If we include the hardware/software backbone necessary for the development of any other modern 'technologies', it becomes clear that nearly all such innovations depend on the advances of electronics and ICT. From the design and planning of large public infrastructure projects like bridges and roads to the assembly of complex chemical compounds and the deployment of new energy sources, ICT permeates every technology development process.

The dominance of ICT/electronics in the global industrial landscape began in the 1970s and shows no sign of decay. The Technology and Innovation Report published by the UN (Fig. 1) shows that ICT-related technologies are a key industry in the recent economic growth of the Global North (Red). It also predicts a persistent growth dominated by similar technologies (called 'Industry 4.0' or 4IR (Marr, 2018)). Simultaneously, the report clearly indicates that within the same period of ICT dominance created a widening gap in wealth distribution between the Global-North and the Global-South (Blue), a gap that is expected to widen even further (for a similar discussion on the root cause of inequality between the Global North and the South see (Naude & Nagler, 2017; Vu et al., 2020). It is not that other industries haven't played a significant role in this context; rather, the pervasiveness of ICT across everything else is the key aspect to note.

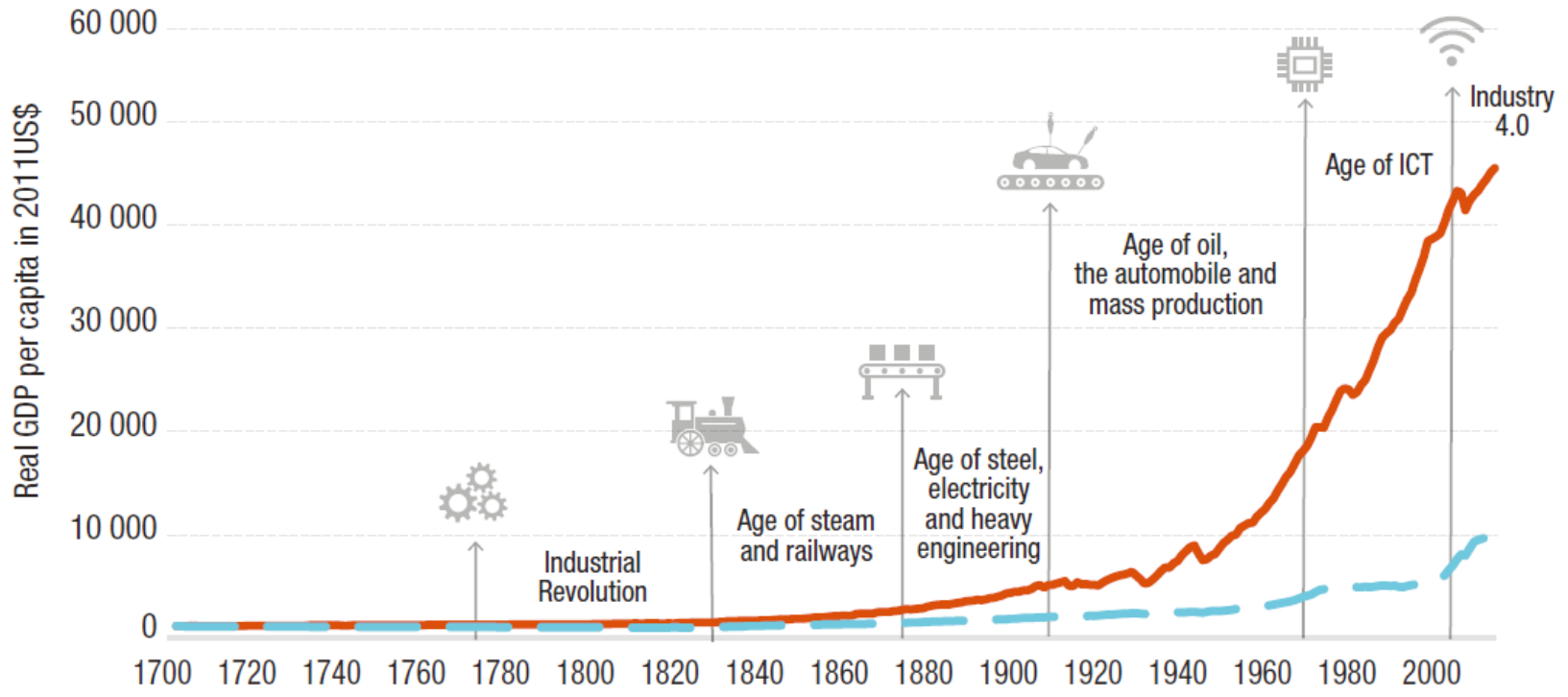

*Figure 1. Adapted from the UN Conference on Trade and Development report (UNCTAD, 2021). Red line represents Global North countries (Western Europe, US, Canada, Australia, Japan, New Zealand) and blue line is the Global South countries (not in the first list).*

Today, all major industries (e.g., pharmaceuticals, automotive, entertainment), are inseparable from their ICT/electronic infrastructure. Industries rely on extensive cloud services not only for managing email and spreadsheets, but also for product/service modelling, supply chain management, data storage/analysis, and AI algorithms. The ICT backbone, which physically represents servers, gadgets and devices (all full of electronics), has now become indispensable to the success of prominent companies across all industrial sectors. Consequently, ICT-related innovations have a direct impact on numerous industries beyond their own.

## 3. Semiconductor microchips as the core of ICT Innovation

Following this argument, we emphasise that electronic technology, primarily semiconductor-based electronics (i.e., microchips), is one of today's most important general-purpose technologies, driving much of modern ICT (and most other) innovation. This includes not only the virtual/cyber world but also numerous everyday physical devices and services, practically anything that can be turned ON/OFF. Microchip innovation has indeed resulted in everything from our ever-powerful, cheaper smartphones to gigantic data centres, from the magic of AI algorithms to ever-precise missiles in modern warfare. Driven by the promise of Moore’s law—the observation that the number of transistors on microchips doubles roughly every two years—, microchips have not only become more powerful in terms of what they can do, but have also massively proliferated in number (McGregor, 2022). This immense importance of semiconductor/electronics was not always clear to the wider public. But the geopolitical climate of the last few years (Ahmed, 2025; Peters, 2022) and sanctions on China have brought the ‘chip war’ into the public domain (Miller, 2022).

Current innovation in semiconductor microchips is the fastest we have seen in any sector in our lifetimes and possibly in human history. Starting from around 2000 components in a microchip in the 1970's, it has reached 50 billion today, a seven-order-of-magnitude increase in ‘efficiency’. This ‘success story’ had a significant positive impact on several aspects of modern life. Beyond enabling a more connected society, ICT innovations have advanced medical/automotive technologies and long-distance communication, and have democratised access to various goods and services over the past few decades. However, unlike any other industrial sector, the semiconductor industry has, for over 30 years, produced a roadmap (ITRS: International Technology Roadmap for Semiconductors) that dictates most research worldwide (IEEE IRDS, 2023). These projections roughly follow Moore’s law and provide a direction in R&D investment for the next 10-15 years. Due to the massive investment in financial and intellectual capital, the industry most often meets its projections, creating a cycle of forecasts for another round. For researchers in industry and academia, these roadmaps are a key driver of innovation, with little assessment or reflection on who requires or benefits from these technologies. But once created, microchips need to be sold in very large volumes to profit from these investments. Hence, new devices, services, and needs are constantly created to justify this cycle of improvements in microchip technology.

Contrary to common understanding, microchips do not only perform computation in digital processors. All kinds of electronic devices benefit from their continuous innovation. These include analogue/photonic devices, integrated sensors/actuators etc. Their progress is often grouped under the More-than-Moore (MtM) trajectory, which doesn't strictly follow Moore's law scaling, with its own ITRS roadmap.

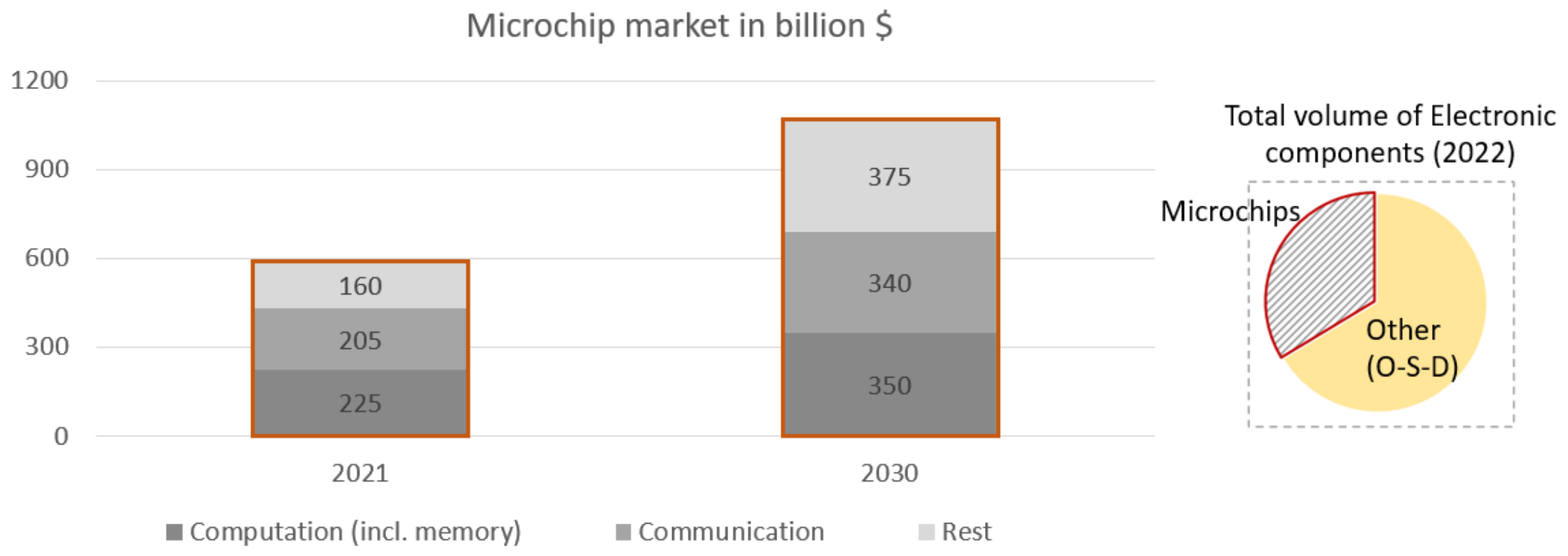

*Figure 2. Global semiconductor microchip market prediction (Burkacky et al., 2022). The 'Rest' row includes every other kind of microchip necessary in various industries. The volume shipment (2022) of all primary electronic components (microchips and non-microchip) is shown on the right.*

Compared to their relative invisibility in our lives, it is often a surprise to realise the size of the global microchip market (nearly as big as that of pharmaceuticals), something that is dramatically increasing with the AI hype (Fig. 2). However, it is interesting to note that a third of microchip market value comes from MtM category (beyond computation and communication). These are low-cost microchips, but their affordability drives innovation and growth across other industries. Then there are even more primary electronic components that are not 'microchips' (often called O-S-D: optical-sensors-discrete), and they form the largest share in volume shipment (inset Fig. 2). However, all of these semiconductors do not emerge from a vacuum, nor do they exist in isolation once they have been produced. Their manufacturing requires extraction and consumption of enormous amounts of material and energy (Crawford, 2022; Pirson et al., 2023).

High-profile AI systems today are made of several thousand cutting-edge computing microchips (along with many more less advanced ones), consuming huge amounts of energy for both computing and cooling. With the growing hype around AI, new generations of these microchips are being built continuously, with no end in sight (Edwards, 2025). However, the 'promise' of the Internet of Things (IoT), which aims to connect everything (from microwaves to factory-farmed cattle) to the web, results in an even larger number of microchip units. The number of IoT devices (each including several semiconductor components) is expected to reach tens of billions of physical objects by the end of the decade (Taylor, 2025). While they are often presented as a promising solution for (energy) efficiency, the impact generated over their life cycle is mostly overlooked until recent cradle-to-grave life-cycle studies (Moreau et al., 2021; Pirson & Bol, 2021). Along with their considerable carbon footprint, we will have all these devices in the environment for years to come (alongside their accumulated obsolete predecessors), adding to e-waste eco-toxicity. In fact, disposable FastTech (vapes to mini-fans) might be a much bigger problem looming on the horizon (over 1billion bought by UK consumers in 2024 alone, and half of them thrown away (Banfield-Nwachi, 2025))

## 4. Innovation in ICT sector and its impact

We have shown that ICT is not confined to a single industry but rather permeates almost all modern industries and technologies. Moreover, in popular discourse, the terms 'technology' and ICT/electronic/semiconductor have become almost synonymous. This interchangeable usage is even observed in multinational consulting firms (e.g., Deloitte, McKinsey), which serve as crucial information sources for investors and policymakers (Mazzucato & Collington, 2024). Though electronics (made of semiconductor microchips) is inherently assumed to be part of ICT, but it often plays the biggest part. In light of this, we propose introducing the term ICET (Information,

Communication and Electronic Technology) to encapsulate this all-encompassing technological domain.

A significant portion of humanity (primarily in the Global North) has become accustomed to witnessing the relentless cycles of innovation in ICET, which have created new computing classes every few years (Fig 3). We employ the term 'cycles' to emphasise that the rapid technological advancements, leading to faster/smaller/cheaper devices and services, necessitate the development of new algorithms/servers/hardware, thereby triggering further innovation in ICET. As a result of this, such innovations have acquired a unique position in the collective consciousness of the general public. But what seems a natural and deterministic development of technology is the result of a distinctive alignment between specific societal values and interests (Pansera & Fressoli, 2021; Winner, 2014). In particular, ICET innovations align with the values of capitalist ventures in Silicon Valley by enabling efficiency, data monetisation, scalability, and global market reach (Wark, 2021). These principles are fundamental to the consumerist ethos prevalent in the Global North (but that are quickly colonising other societies, e.g, China/India), driving the success of these industries (Vinsel & Russell, 2020).

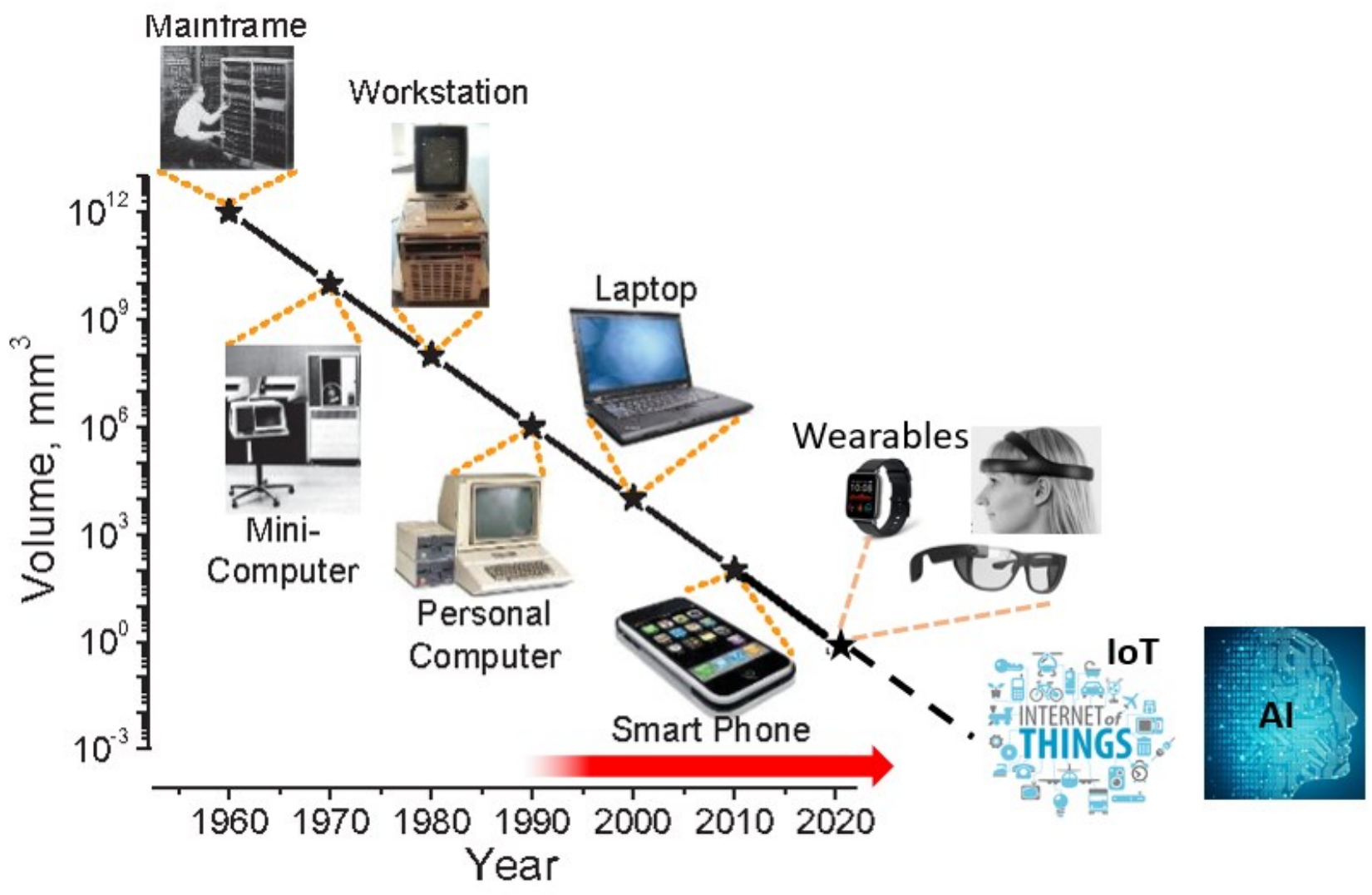

*Figure 3. Bell's law of new 'computing class', roughly every decade. New classes are smaller, more 'efficient' and create new markets (adapted from (Lee et al., 2012)).*

It comes as no surprise that the largest, most profitable, and globally renowned companies in existence today, such as the 'magnificent seven'—Alphabet, Amazon, Apple, Meta, Microsoft, Nvidia, Tesla—are predominantly ICET companies. The leaders of these companies harbour a vision of a future in which technological solutions play a pivotal role in addressing the world's most pressing challenges, including those arising from tech itself. This techno-solutionist narrative, which centres on the belief that technology can solve complex societal issues, extends beyond the tech industry. It has gained a strong influence over public opinion and even philanthropic funding (Haven & Boyd, 2021). The allure of technological advancements as the sole problem-solving tool has permeated sectors ranging from healthcare and education to environmental conservation and transportation. Governments, organisations, and individuals alike are drawn to the promise of transformative ICET solutions that can revolutionise the way we live, work, and interact with the world. However, it is important to approach this techno-solutionist narrative with a critical lens. Balancing the benefits of technological progress with responsible, sustainable practices is imperative to ensure that ICET-driven developments align with society's broader interests (Heilinger et al., 2024; Sætra, 2021). Though these technologies have had great positive effects for many, however, not everyone's life has been equally improved, especially in the Global South. Tech enthusiasts might also want to ignore that the innovations of the past two decades have hardly made any dent in the inequalities in the Global North itself (Antonietti et al., 2025). Income inequality, extremely low social mobility, unequal access to health care, underrepresentation of women across tech sectors, and other social inequalities are consistent issues within OECD countries, and haven't shown major improvement in recent years. In our technologically affluent world, even childhood poverty has remained stagnant at around 20% in the UK and in the US for many years (UNICEF, 2025). Digital inequality is prevalent in wealthy countries awash with technology, and has had severe consequences during the COVID-19 lockdowns (Watts, 2020).

In contrast to what it can and did actually achieve, the widespread optimism and belief in tech innovation (particularly prevalent among technocrats) has not stopped. Considering the amount of financial and intellectual capital that gets poured into this sector, the belief might have even increased. Based on a more comprehensive understanding of ICET and the context provided, we can identify three distinct categories of impact arising from the ongoing innovation cycles: *embodied impacts, induced impacts, and implied impacts*. While a three-tier environmental impact of ICT has previously been established by Hilty (Hilty & Aebischer, 2015), our focus is not on the impacts of individual devices, but rather on the broader consequences arising from the paradigm of continuous tech/ICET *innovation*. By identifying these three impact categories, we aim to capture the multifaceted effects arising from the interplay between these innovations and various sectors of society. Embodied impacts refer to the obvious tangible consequences directly associated with this innovation cycle. It is the result of production, consumption, and disposal of new technologies (along with the old). Induced impacts encompass the ripple effects triggered by ICET advancements throughout the economy, such as increased embodied impacts from other sectors, including changes in employment patterns, business models, and societal behaviours. Lastly, implied impacts consist of techno-solutionist beliefs and narratives leading to the promotion of over-optimistic expectations about future developments of ICET. By exploring these impacts, we aim to provide an understanding of the wide-ranging effects arising from the relentless cycle of innovation.

## 4.1. Embodied Impact: Material Consequences

The *embodied* impact represents one of the most noticeable aspects of the innovation cycle in ICET sectors. It encompasses factors such as increased carbon emissions, water usage and other environmental footprints. While it is also important to acknowledge that these innovations have well-documented societal impacts, such as surveillance capitalism (Zuboff, 2019) and ethnic discrimination (Benjamin, 2019), they fall outside the scope of this discussion.

The assessment of environmental impact, particularly in terms of direct carbon emissions of data centres, has received increased attention in recent studies. However, it is crucial to consider the broader ecological consequences, including the significant impact of ICET production and its e-waste pollution. Due to the complex nature of the problem, accurate numbers on $CO_2$ emission of ICET have to be constantly revised and updated (Gupta et al., 2022). Significant uncertainty remains (O'Brien, 2024), but one of the most comprehensive studies estimates the global warming potential (GWP) of ICT at around 4% (Freitag et al., 2021), which is already higher than that of the aviation sector. However, the comprehensive analysis of water/energy consumption for just microchip production is still in its early stages (Moreau et al., 2021; Pirson et al., 2023). The materiality and associated eco-toxicity of the production phase are also expected to result in a much larger environmental footprint than previously assumed (Roussilhe et al., 2025).

Furthermore, these numbers are based on the current rate of tech development. New cycles of ICET innovation (including the AI hype) can dramatically exacerbate the problem. Due to continuous shrinkage of semiconductor microchips, according to Bell's law, a new computing class is introduced roughly every decade (Fig. 3) (Bell, 2008). This creates new markets and applications, but older devices/technologies also remain in the market and continue to accumulate. It's notable that most companies in this sector not only continue producing older-generation microchips (first released over 30 years ago), but also do so at the same peak volume. Similarly, older computing classes remain in use for many years (red arrow in Fig. 3) even when new ones are created. Hence, ICET innovation results in the cumulative addition of many more tech products on top of what we already have.

ICET innovation is also destructive in nature, particularly from end-user perspective. Older devices/services need to be discarded (or become unusable) at an alarming rate. This entire economic model is built on making older technology obsolete by either overloading them (e.g., with updates) or making repair impossible. The idea of planned obsolescence is built within the ICET innovation paradigm and is a prime driver for profit (Hadhazy, 2016). Apart from its environmental impact, the rate of obsolescence (both planned and unplanned) creates an unprecedented amount of legacy technology that often needs to be kept alive for decades. With declining official support (from manufacturers), maintenance is often done by volunteers or via a risky, illegal market. This mostly impacts vulnerable people, organisations (e.g., schools, libraries, small businesses), and cash-strapped public sectors (e.g., hospitals, local transport) that are locked into legacy systems, e.g., older Windows versions (Chaette, 2020). For a large part of society, these upgrades would require significant infrastructural investment at regular intervals, which is unaffordable.

## 4.2. Induced Impact: Enabling Effects

The *induced* impact can be defined as the increased dependency on ICET from other sectors that prompts/enables further innovation cycles. As indicated earlier, many non-ICET industries can create new goods and services only because of the availability of this general-purpose backbone technology. New material development, advanced processing plants, food/agricultural technology, and supply chain management are all affected by (and often highly dependent upon) ICET innovation. ICET is generally credited with increasing efficiency in many sectors, but greater efficiency is almost always associated with higher production (due to the well-known 'rebound effect' (Galvin, 2015)), nullifying the advantages gained from individual efficiency.

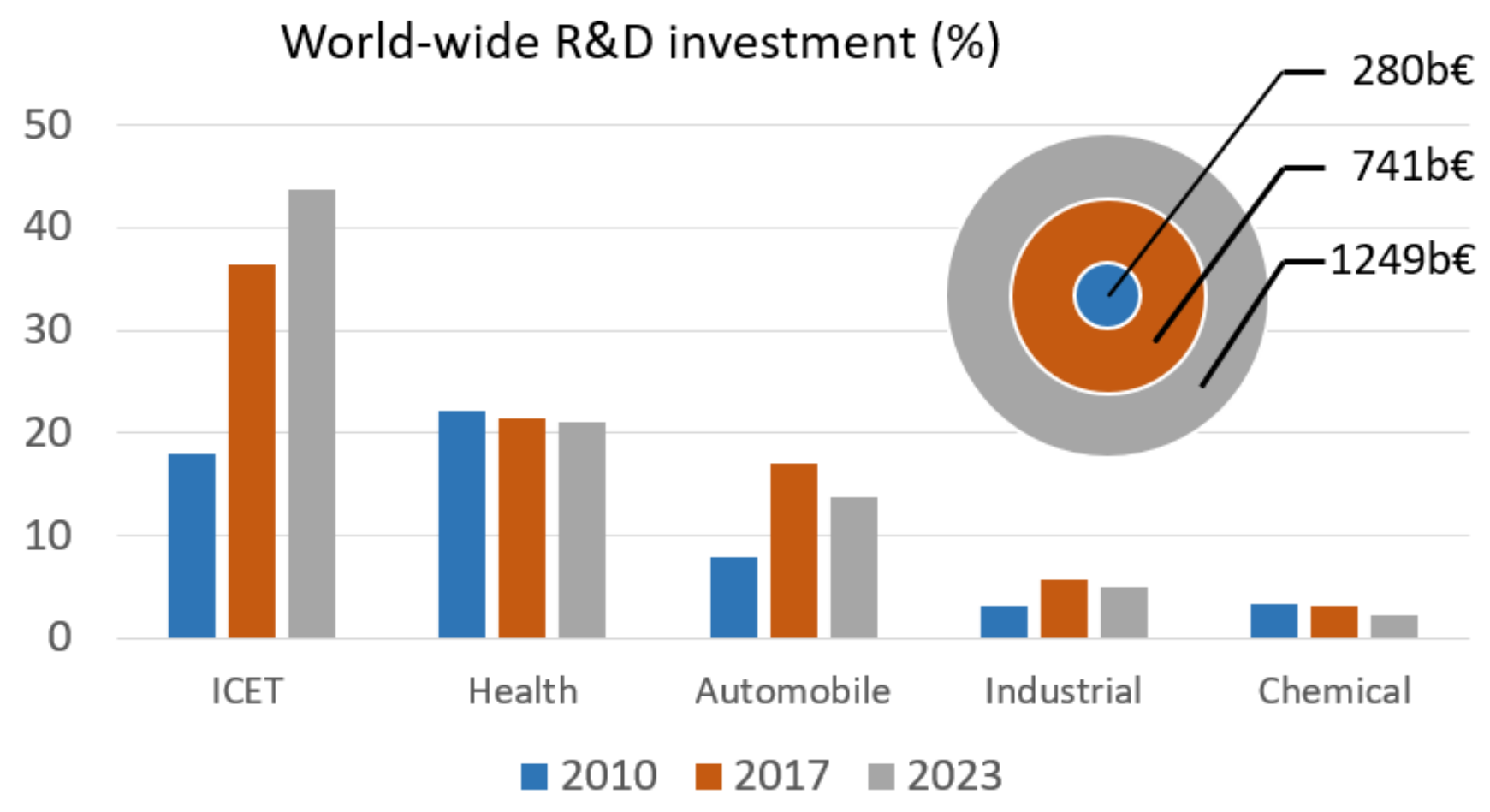

*Figure 4. Percentage of worldwide R&D investment in various sectors and the total R&D budget (inset). Source: EC Yearly Investment Scoreboard (*Economics of Industrial Research and Innovation*, 2025)*

For example, automotive and aviation, two highly ICET-depended sectors, have advanced hugely over the past few decades. This has required intensive modelling (e.g., in fluid dynamics), new material development (for lighter chassis), several connected sensors and of course a highly automated production line. Very little of this would be possible without advanced ICET (Armstrong et al., 2020; Cimmino et al., 2020). This has resulted in increased efficiency (both in fuel consumption and cost) in all types of cars, including electric vehicles. However, gas-guzzling SUVs have also become more attractive due to several such new features and had a dramatic increase in market share in the same period (From 15% to >50% in the last 15 years (Cazzola et al., 2023)). Similarly, along with the efficiency improvement in aviation fuel consumption, more flight miles are taken globally than ever before (Duggal & Haddad, 2021). The material footprint of the extractivism and end-of-life pollution of this affluence is rarely borne by the population in Global North (or the comfortable upper-class in Global South), who benefits from the technology (Hickel et al., 2022).

The innovation rush in this sector can be clearly observed in the worldwide investment portfolio across various industries (Fig. 4), compiled periodically by the EU (Grassano, et al., 2022). The R&D investment in total ICET (under two items, 'ICT services' and 'ICT producers') was already significant in the last decade, but it is now comparable to the next 4 sectors combined. This happened at the same time as the total R&D investment budget increased significantly as well (inset). Much of the innovation in the next two sectors (healthcare and automobile) is also a product of accompanying advances in ICET. This massive expenditure and resulting product cycle come with an environmental cost that is far greater than the embodied impact of ICET alone. The scale of investment and its unparalleled dominance over all other sectors also created the need for greater ICET innovation to justify the financial risks taken in the first place.

## 4.3. Implied Impact: Ideological Constructs

Finally, we define *implied* impact as the effect that ICET innovations have on the construction of over-optimistic expectations about the future of technology and the benefits it can bring. We argue here that the more than five-decade-long exponential growth of this sector and its pervasive nature in virtually all modern technological achievements play a crucial role in reinforcing a narrative of ineluctable, endless progress. A scenario in which most of the problems that humanity faces—from climate change to inequality and poverty—can be solved with technology. This vision of the future is often associated with Silicon Valley-style entrepreneurs and tech-visionaries who advocate making everything 'smart' and 'connected'. They expect all of these to be secure, profitable, and also provide eternal economic growth for the foreseeable future (Andreessen, 2023; McAfee, 2019; Mills, 2021). The future positive impact of technological innovation on most people's lives is implied, whereas the immediate harm on the environment and marginalised communities are often neglected or forgotten (Jacobson, 2023).

For example, entire economies and businesses operate within a conceptual framework that assumes mobile phones will become faster and hold more data. This is not restricted to just consumer electronics. Research into other technologies (including ICET) is highly influenced by the same expectation bias (Lombardo et al., 2022; Rodriguez et al., 2018). It is assumed that if more data could be collected, more modelling could be done, and more complex algorithms could be deployed across the board, most problems (including UN Sustainable Development Goals) can be solved (Vinuesa et al., 2020). This *implied* impact shapes and reinforces collective expectations about the ever-increasing ability of technological solutions to fix any problem that could arise in the future (Gates, 2021; Jacobson, 2023). These expectations are often framed in terms of technological determinism (i.e., technological progress as 'inevitable and unstoppable') (Pansera & Fressoli, 2021), eco-modernism, techno-optimism or techno-solutionism (Asafu-Adjaye et al., 2015; McAfee, 2019). An example of this myth is the notion of green growth, which relies on the belief that economic growth can be sustained *ad infinitum* and decoupled from its environmental impacts through massive investments in new technologies (Pollex & Lenschow, 2018).

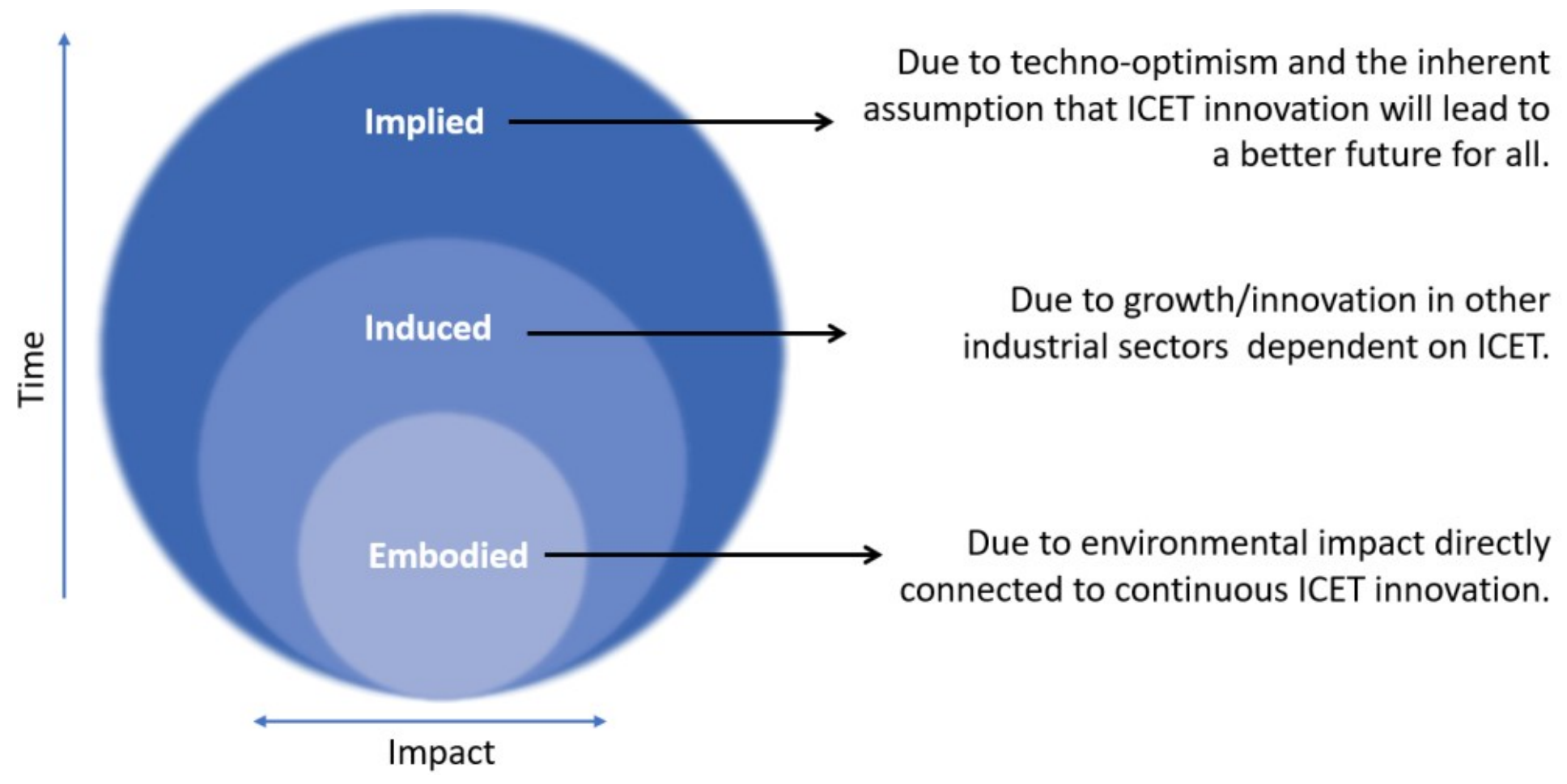

*Figure 5. A qualitative visualisation of impacts from continuous ICET innovation. The embodied, induced and implied impacts of a ICET innovation can start at the same time, but could have different sizes and could extend affect the society in different ways over time.*

Fig. 5 shows a qualitative diagram representing these three kinds of impacts. While they can be experienced simultaneously when a new technology develops, their impact sizes will vary. Most research has focused on identifying the *embodied* impact of ICET, but it remains difficult to measure. Since the *induced* impact exacerbates environmental damage via many more industrial sectors that are not considered part of ICET (from fast-fashion to fossil-fuel), its size can be considered much larger than *embodied* impact. All the *induced* environmental impacts will also take longer to show up. However, above all, the *implied* impact is possibly the largest of these, and will stretch much farther into the future. It results in an ideology and a set of justificatory narratives that drives long-term policies and investments. We further analyse this ideology in the next section.

# 5. Exposing ICT-Based Techno-Solutionism

Having explored some key aspects of ICT deployment and development, we see why proposed technological solutions to societal problems often lead to undesirable outcomes. One reason is that human foresight capabilities are far from the level needed to predict and forecast, with any precision, the consequences of deploying new technologies in complex socio-technical systems. Another reason, and the one we mainly focus on, is that any solution based on ICET will have significant negative impacts on both people and the planet, and these costs are poorly understood and routinely overlooked by proponents of technological solutions. In the following, we discuss the main factors to consider when evaluating techno-solutionist proposals for social problems.

## 5.1. The Myth of Growth-Based Solutions

The first factor driving techno-solutionist approaches is our ongoing expectation of GDP-based economic growth. Continuous economic growth has been a leading policy expectation of capitalist economists (and also in socialist countries) since the WWII. There have been several critical analysis of this idea its long-term, multifaceted impact on humanity and the planet (Kallis, 2017). While limiting our carbon emissions has become a stated goal for many countries and industries, the ongoing push for GDP growth and its associated carbon emissions continues to rise. Though some countries have managed to restrict their territorial carbon emission (by offshoring production and disposal), there is no viable way to reduce worldwide emissions within the current growth paradigm (Hickel, 2021). In contrast, emissions have continued to increase (in line with GDP growth) even after the Paris Agreement, contrary to the nearly 7% per year reduction necessary. Both Neoclassical and Evolutionary economic models operate under the assumption that technological innovation drives GDP growth, which in turn leads to job creation, increased welfare, and overall prosperity. This model also suggests that technological solutions can effectively address the challenges that arise along the way (Pollex & Lenschow, 2018). Nevertheless, it is important to note that the research literature on Responsible Research and Innovation has shown that technical progress is more likely to entail unforeseen risks and generate winners and losers rather than universally beneficial outcomes (Stilgoe et al., 2013).

More recently, the pursuit of endless economic growth has been criticised by de-growth/post-growth scholars. This heterogeneous field of study questions the environmental feasibility of infinite growth on a finite planet and also the unsustainability of consumerist societies (Hickel, 2021; Paulson et al., 2020). As we have indicated before, ICET is often the most prominent of all technologies discussed in the context of growth (Fig. 1). However, post-growth and de-growth scholars still do not have a common position on the role of ICET innovation in such a transitional economy (March, 2018; Robra et al., 2023). The potential of tech innovation to decouple our energy demand from GDP growth, whether in relative or absolute terms, has been subject to scrutiny (Lange et al., 2020). Some authors have even gone so far as to argue that the pursuit of techno-solutionism, along with the techno-utopian dream, has become an end in itself, verging on fetishisation (Kerschner et al., 2018). These critical perspectives challenge the simplistic narrative that technological innovation alone can solve all societal and environmental challenges.

We push this argument further to suggest that the firm belief and expectation of perpetual ICET innovations are a dominant factor in this particular utopian narrative. We believe that this is not only necessary but that it also makes the point stronger and not touched by other similar work (Pollex & Lenschow, 2018). There is little doubt that the iPhone is a great success story in human ingenuity. But we didn’t stop with just one (or even ten) versions. Similar ICET advances contributed to our perception that ‘tech’ could deliver almost magical solutions at an unimaginable rate. Techno-utopia is the reason for our hope in many recent innovations in ‘green’ technologies. For example, electric cars, solar cells, wind turbines, smart grids, carbon capture and storage (CCS) are all proposed technological solutions to current environmental problems. Though they have a role to play, all of these require further technological consumption and a massive ICET-supported infrastructure. None of these new devices come out of thin air or vanish on their own. Research has shown that building these technologies and services at a global scale would require unprecedented resource extraction, especially from the most vulnerable populations (Martínez-Alier, 2012). Environment cost of lithium mining for EV batteries (Nast, 2018), water usage and GHG emission from semiconductor manufacturing (Pirson et al., 2023), impact of rare earth metal extraction for various 'green' technologies (Pitron, 2021), or environmental cost of scaling futuristic CCS (Bhowmik & Grant, 2022) are now well documented. Though we have become more aware of ICET's environmental footprint in the last years, it still appears that even researchers working on climate change can be prey to the lure of techno-solutionism.

## 5.2. The hidden ICET

Another reason to be wary of ICET-based techno-solutionism is what we call the "hidden ICET". The fact that one of the biggest ICET users—and funders—are intelligence services and the military often goes without careful scrutiny. Various ICET innovations are direct consequences of what the military/security services want in their own ideal world. Most of the original microchip development occurred due to the US military's need for more accurate missiles (Miller, 2022), and similarly, the earliest major backer of large-scale AI also came from the military (Konaev et al., 2020). But one country's security is a threat to another, even within the intelligence alliances of the Western world (Švenčionis, 2022). China has also shown the world the level of data gathering and surveillance it can do on its own people (Yeung, 2025). Setting aside what good (or extreme harm) this is doing, the question of the environmental cost of such ICET infrastructure remains (Crawford, 2022). This is the hidden ICET, where enormous defence budgets depend on sustained innovation for no particular end other than projecting the state's power over potential external (and internal) enemies. In fact, the recent geopolitical tensions and the current US administration have encouraged the largest tech companies to entirely drop their reluctance to develop military-ICET (Corbyn, 2025). New start-ups and funding has generated rapid innovation in these areas, that will spill over into civilian technology, creating more markets for devices/services/gadgets in an endless cycle of consumerism.

The highly speculative financial industry is another example of excessive hidden dependency and the drive for ICET innovation. Massive infrastructure is built, and algorithms are created for high-speed trading to gain a millionth of a second advantage in data latency (Baker & Gruley, 2019). High-frequency traders and hedge funds generate unimaginable revenue with every fraction of a second improvement in computing resources (Kerner, 2021). Dedicated tech infrastructure for bitcoin mining was one of the reasons for its exponential growth and unsustainable environmental footprint (Sial et al., 2025).

## 5.3. Who benefits from ICET innovation?

An important factor explaining why ICT-based techno-solutionism is harmful is that the benefits and rewards from this development are deeply unequal (D`ignazio & Klein, 2020). Correspondingly, the interest of upholding the status quo is strongest among those who benefit most, even if this comes at the expense of others and the environment. It is no surprise that, when the world was struggling to deal with the COVID-19 crisis and lives and livelihoods were lost in unprecedented numbers, a certain group of people benefited enormously. These people who accumulated the most wealth during the pandemic years were primarily ICET company owners and shareholders (BBC, 2021). The hypothesis that they deserve it, since such technologies were key to keeping society running, has to be understood in context. It can be argued that the omnipresence of remote working among white collar workers has kept the high-earning section of the community completely oblivious to the life-and-death struggle of people who had to show up for work physically. The high socio-economic disparity in mortality, morbidity, and educational outcomes after COVID is a lasting testimony to how our dependence on ICET solutions protected only some people (Golden et al., 2023; McGowan & Bambra, 2022).

Unsurprisingly, there was hardly any dent in the cycles of superfluous innovation in these companies, even during this global crisis. New phones and gadgets were launched with the same regularity and fanfare. It is often obvious that these innovations do not have any intrinsic value other than helping these massive profit-making businesses become even bigger. The hype in AI has truly taken this to yet another level, and nothing represents it more than the sheer audacity of Open-AI CEO courting Saudi billionaires for a 7-trillion dollar investment in microchips and other AI infrastructure (Fitch, 2024) – a sum larger than the entire US federal budget ($6.3t in 2022). While the industry keeps

growing at a rapid rate with an alarming amount of resource consumption, there is scant evidence of AI's true benefits. Even the long-promised AI-based drug discovery has, in fact, shown very little progress (Kuchler & Heikkilä, 2025). Nevertheless, huge profits will be rolled out to investors and technocrats, as they all promise to solve humanity's 'wicked problems' using AI and all things tech (Amodei, 2024; Ferres & Weeks, 2024). As shown in Fig. 6, the world's largest R&D spenders are all ICET companies, with mammoth budgets compared to public funding agencies (which have to fund all science and humanities domains). With such disparity, there is no doubt what kind of innovation dominates the public imagination (Mitra & Raskin, 2023).

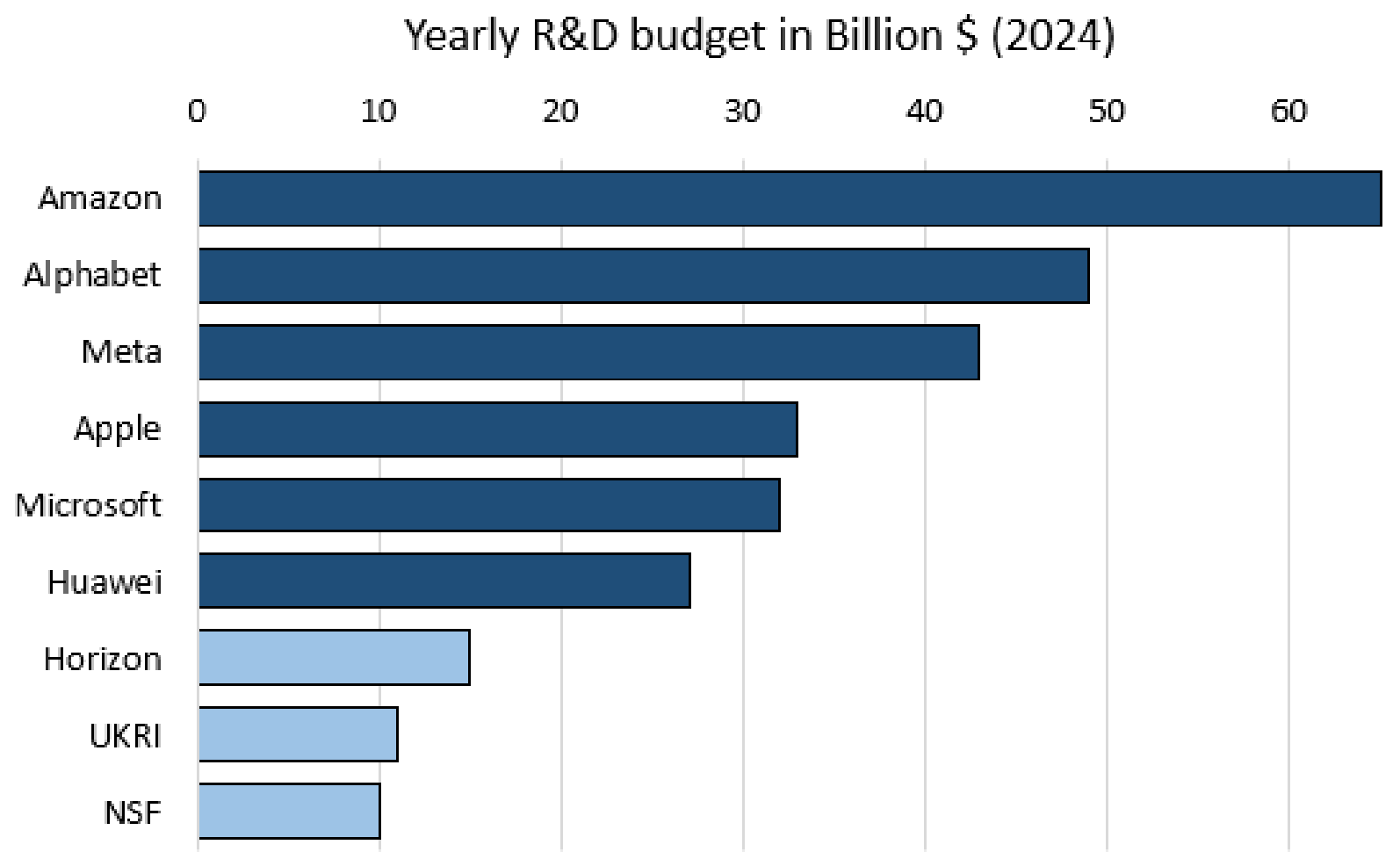

*Figure 6. World's top six R&D spenders (dark blue) compared with some of the major public funders (light blue). Due to limited data, Amazon's 2024 budget is an estimate. (Galan, 2025; Mitra & Raskin, 2023)*

Popular magazines, websites, YouTube channels etc. testify to a lasting enthusiasm for tech innovation. However, little is discussed about the lax regulatory oversight that has always enabled these innovations to thrive. Some of the largest semiconductor industry players' proposed GHG-reduction plans fall well short of the necessary environmental targets (Gockce & Mena, 2022). That, too, with the assumption that best practices will be followed everywhere and that there will be no new growth areas. The true cost of ICET innovation would have been much higher if all the environmental and societal impacts had been considered. As ICET companies are among the largest lobbying groups in both the US and the EU (Birnbuam, 2023; Max et al., 2022), it is unlikely that they will pay this cost anytime soon.

One surprising fact in this industry over the last two decades has been Taiwan's clear dominance in producing the most advanced microchips (e.g. ones designed by Nvidia) (Buchholz, 2023). However, the recent geopolitical climate has completely disrupted this long-standing status quo. With unprecedented speed, the US and the EU have pumped in Cold War-era investment in semiconductor facilities ($280 billion through the US CHIPS Act and €43 billion through the EU Chips Act, both announced in 2022). This will no doubt create more competition among players in the ICET industry, inevitably resulting in greater pressure to innovate in ways that feed into things we do not need and the planet cannot afford.

## 6. Conclusion: Reconsidering the value of ICT innovation

We argue that the tremendous 'success' in ICET innovation in the last five decades not only has an enormous *embodied* and *induced* impact on the environment (and society), but the narrative it perpetuated for so long has created a new kind of *implied* impact: It has cemented the idea that technology alone can create prosperity for all and solve all of humanity's problems. While innovation in some basic infrastructural technology (e.g., sanitation, building material, public health, agriculture) certainly has a high potential to uplift the quality of life in many parts of the world, some of the most important progress in the human condition has very little to do with recent technological innovation (Kitcher, 2021). Universal human rights, universal franchise, the right to education, the 5-day work week, 8-hr workdays, minimum wage, pensions, workers' unions, universal healthcare (in some countries), universal basic income (hopefully), etc., are all progressive social innovations promoted by direct political action and mobilisations. Furthermore, what makes our societies work is the maintenance and care of not only existing technologies but also of people, the environment, and non-human entities. Techno-utopia and techno-solutionism are myths sold to us to defer all possible problems until the magic bullet arrives (invariably with its own problems). ICET innovation is not only complicit in creating and perpetuating this narrative, but it is also the technology sector that stands to gain the most from this dangerous ideology.

This paper did not aspire to identify alternative models of tech/ICET innovation that could be beneficial to society. However, it should have become clear that the value of existing innovation cycles are ambiguous and stands in need of reassessment. Given that, as we have shown, constant "innovation" and growth perpetuate or even aggravate problems such as environmental degradation and inequality, we believe in the need to limit ICET innovation/growth and explore prospects of de-growth/post-growth also in this sector. It will be a task for future research among like-minded scholars to formulate practical pathways in this direction (Pansera & Fressoli, 2021). Steps could include Extended Producer Responsibility (EPR) as a cost integrated into ICET (Gu et al., 2019), a novel way to reimagine 'innovation' beyond tech specs that include social/environmental metrics of progress, and a general balancing of the power and interests of those already better off against the needs and interests of those worse off (Heilinger, 2024). Similar to other technologies that can cause harm (e.g., nuclear science and virology), even some limits could be established on what ICET research/innovation can and cannot be funded. But unfortunately, the EU's overarching policy package 'Fit for the Digital Age' is a call for building more ICET, and is often in contradiction of it's own goal of 'European Green Deal' (Santarius & Lange, 2022). The critique proposed in this paper, and its possible policy implications, are contrary to the current *zeitgeist* and will be met with skepticism, even resistance.

Yet, scholars and digital activists have recently proposed interesting analyses and frameworks that resonate with a degrowth-compatible conception of ICET and complement the contributions of low-tech movements (Tanguy et al., 2023). Concepts like slow computing (Kitchin & Fraser, 2020), post-automation (Smith & Fressoli, 2021), technopolitics and platform cooperativism (Scholz, 2016) are promising critical paradigms that explore potential alternatives to the ways capitalist societies frame the role of ICET in society. However, the scale of ICET infrastructure, along with its contradictory indispensable/invisible presence in everyday life, is often not emphasised enough. With the present work we have tried to elaborate on this issue and call for new scholarship that reconciles the demand for legitimate ICET innovation with that of a just, equitable planet that does not exceed its boundaries.

## 7. Funding

The authors did not receive any funding for this work.